# A determination of the pairing interaction in the high $T_c$ cuprate superconductor $Tl_2Ba_2CaCu_2O_8$ (Tl2212)

W. A. Little[a] *, M. J. Holcomb[a], G. Ghiringhelli[b], L. Braicovich[b], C. Dallera[b], A. Piazzalunga[b], A. Tagliaferri[b], N.B. Brookes[c]

[a]*Physics Department, Stanford University, Stanford, CA 94305, USA*
[b]*Politecnico di Milano, Milano, Italy*
[c]*European Synchrotron Radiation Facility, BP 220, F-38043, Grenoble, France*

**Abstract**

We have measured the near-normal reflectance of $Tl_2Ba_2CaCu_2O_8$ (Tl2212) for energies from 0.1 to 4.0 eV at room temperature and used a Kramers-Kronig analysis to find the complex, frequency dependent dielectric function, $\varepsilon(\omega)$ from which the optical conductivity $\sigma(\omega)$ was determined. Using Thermal-Difference-Reflectance (TDR) Spectroscopy the reflectance of the sample in the normal state just above the superconducting transition, and in the superconducting state were then obtained. From these data we determined the ratio of the superconducting- to normal-state optical conductivities, $\sigma_S(\omega)/\sigma_N(\omega)$. Mattis and Bardeen had calculated this function within the BCS theory, where the gap is a fixed energy-independent quantity. Taking into account the retarded nature of the electron-phonon coupling results in a complex, energy dependent gap $\Delta(\omega)$ causing deviations from the Mattis-Bardeen plot at energies where the phonon coupling function $\alpha^2(\omega)F(\omega)$, is large. We find a typical deviation near the phonon energies in Tl2212, and in addition, at 1.2 and 1.7eV. The phonon, and these electronic terms can each be described by a coupling constant $\lambda_i$. None of which by itself gives rise to a high transition temperature, but the combination does. Using Resonant Inelastic X-Ray Scattering (RIXS) we find that the d-to-d excitations of the cuprate ion in Tl2212 fall at the same energies as the dips in the Mattis-Bardeen plot. We conclude that the high superconducting transition temperature of the cuprates is due to the sum of the phonon interaction, and interactions with the Cu-ion d-shell.

*Keywords: RIXS, superconductivity; cuprates; optical conductivity; electron-phonon interaction; exciton interaction;TDR Spectroscopy;*

## 1. Introduction

The understanding of the mechanism responsible for the high superconducting transition temperature of the cuprate superconductors has remained elusive in spite of the enormous effort devoted to its study during the past twenty years. The materials have a rich phase diagram involving many properties, some of which are possibly unrelated to the superconductivity itself. This tends to confuse the analysis of the cuprates' most distinctive feature, their extraordinarily high transition temperature. The challenge has been to separate the wheat from the chaff. Much has been achieved along these lines by the development of new techniques and improvement of older techniques, in particular, of angle-resolved photoemission spectroscopy (ARPES)[1][2][3][4], scanning tunneling spectroscopy [5], and phase-sensitive, edge-junction tunneling [6]. These have revealed the nature of the Fermi surface, and of the charge carriers, their coherence properties, and clues to the understanding of the mechanism, itself. We report here the results of two other relatively new techniques, one optical, and the other Resonant Inelastic X-Ray Scattering (RIXS) that now provide compelling evidence of the nature of the interactions that cause the high transition temperature of the optimally doped cuprates.

## 2. Background

In the original work by Bardeen, Cooper and Schrieffer (BCS) [7] the pairing interaction due to the phonons was modeled as a constant attractive potential between electrons located within the Debye energy of the Fermi surface. This led to an energy independent gap function, Δ in this region.

Eliashberg [8] extended this treatment to strong coupling taking into account the retarded nature of the electron-phonon interaction. The gap then became a complex, energy-dependent function, $\Delta(\omega)$, whose structure reflected the energy dependence of the electron-phonon interaction, $\alpha^2(\omega)F(\omega)$, which involves the phonon density of states, $F(\omega)$, and the electron-phonon coupling term, $\alpha(\omega)$. The electron density of states then became:

$$N_S(\omega) = N_N \operatorname{Re}\left(\frac{\omega}{\sqrt{\omega^2 - \Delta(\omega)^2}}\right) \qquad (1)$$

McMillan and Rowell [9] showed that this density of states could be measured in tunneling experiments, and $\alpha^2(\omega)F(\omega)$, and the Coulomb pseudopotential, $\mu^*$ determined from it. The energies of the transverse and longitudinal peaks in $\alpha^2(\omega)F(\omega)$ for Pb were found to be in reasonable agreement with neutron measurements, and corresponding critical points to be in excellent agreement with those in the phonon spectrum. *Quantitative* agreement was found between the calculated and observed critical field and transition temperature. The work was extended to In, Sn, Mg and PbIn alloys. It provided incontrovertible proof that the conventional metallic superconductors derived their superconductivity from the phonon interaction alone.

We have completed an analogous series of experiments on the optimally doped high transition temperature cuprates. The high transition temperatures suggested that excitations with energies above those of the phonons would be involved [10][11][12]. If this were so, tunneling would be unable to yield the density of states at these energies, because the voltages required would exceed the breakdown potential of the barrier [13]. However, it has long been known that similar information from that of tunneling can be obtained from the optical properties [14] [15] of a superconductor. At these higher energies, the effect on these properties is small, $\sim\Delta^2/\omega^2$, or about 0.01% of the total reflectance at electron volt energies, so a sensitive technique was needed. We developed such a technique called Thermal Difference Reflectance spectroscopy (TDR) [16]. At a given photon energy, the reflectivity of a sample is measured at two precisely set temperatures, and the ratio of the *difference* determined relative to their *sum*. The instrumental properties cancel in the ratio and the background noise of the instrument could be reduced to about 0.005% of the observed reflectance.

We have used this technique to measure the ratio of the reflectance ($R_S/R_N$) in the superconducting and normal states of six of the optimally doped cuprates [17] [18 [19] [20]. The reflectance ratio was modeled by calculating $\Delta(\omega)$ using the Eliashberg equations [8][21] with a generalized trial function, $G(\omega)$ in place of the electron-phonon interaction, $\alpha^2(\omega)F(\omega)$ and then using $\Delta(\omega)$ to calculate the reflectance ratio from the ratio of the optical conductivity in the superconducting and normal states derived by Shaw and Swihart [15]. The trial function was adjusted until a fit between the calculated and observed reflectance ratio was obtained. Good to excellent agreement was found between the two. In every case, a phonon-electron interaction was needed as well as interactions at energies between 1.6 and 2.3eV.

Several important conclusions can be drawn from these results. First, the optical data shows that the phonon interaction provides a coupling parameter, $\lambda$ of the order of 0.9 to 1.0 in all the materials studied. This interaction alone, while moderately strong, is much too weak to explain the high transition temperature of the materials.

Second, there are no contributions to the pairing in the "mid-infrared region" between 0.1 and 1.0eV that might be expected from some magnetic or plasmon interactions.

Third, the key feature in explaining the high transition temperature is the combination of a moderately strong phonon interaction and the high energy electronic contributions (22). Neither one alone, results in a high transition temperature nor is able to account for the optical properties. Both are needed.

Fourth, the fit to the optical data shows that the real part of the superconducting gap does not remain negative above the phonon energies as it does in superconductors with only a phonon interaction [18][21]. This is a clear indication of an attractive contribution from higher energy interactions that exceeds the repulsive Coulomb interaction at these energies.

Fifth, the real and imaginary parts of the gap function give a characteristic shape to the energy dependence of the calculated reflectance ratio. This is accurately mapped in the observed reflectance ratio both in the phonon and in the electronic-interaction regions. This fingerprint is strong evidence that the Eliashberg equations accurately describe the superconducting gap function even at energies that are many times higher than the phonon energies for which the theory was developed.

In our earlier work, approximate literature values were used for the normal state scattering rate, the screened plasma frequency and the high frequency dielectric function used to model the normal state optical properties. We have since eliminated the need for these assumptions by determining directly the normal state properties from reflectance measurements [23].

## 3. Optical Properties of Tl2212

As reported elsewhere [24][25], we have made a direct determination of the near-normal reflectance of a high quality thin film sample of $Tl_2Ba_2CaCu_2O_8$ (Tl2212) at room temperature over energies from 0.1 to 6.0eV. A Kramers-Kronig analysis was used to determine the real and imaginary parts of the frequency-dependent dielectric function $\varepsilon(\omega)$ from which the optical conductivity $\sigma(\omega)$ was determined. We have refined the technique further with special focus on the region between 0.1eV and 4.0eV. Using TDR spectroscopy in a step-wise manner we determined the reflectance of the sample in the normal state just above the superconducting transition, and in the superconducting state. From these data we extracted the ratio of the superconducting- to normal-state optical conductivities, $Re[\sigma_S(\omega)/\sigma_N(\omega)]$.

Mattis and Bardeen [26] had calculated this ratio within the framework of the BCS theory where the gap is independent of energy. Nam [14] extended this to strong coupling, and Shaw and Swihart [15] showed this ratio could be expressed as an integral over the convolution of the density of states above and below the Fermi energy plus the corresponding case II BCS coherence factors. At high energies the integral could be approximated to:

$$Re\left(\frac{\sigma_S(\omega)}{\sigma_N(\omega)}\right) \approx 1 - \frac{2\Delta_0 Re[\Delta(\omega - \Delta_0)]}{\omega^2} \ln\left[\frac{2\omega}{\Delta_0}\right] \qquad (2),$$

where $\Delta_0$ is the gap at the gap edge. Equation (2) shows a reduction occurs in the above ratio where the gap function $\Delta(\omega)$ becomes large, as in the neighborhood of the electronic terms between 1 and 2 eV for many of the cuprates. Our data shows such depletion. Molegraaf *et al.* also observed such depletion in the oscillator strength sum rule [27] for $Bi_2Sr_2CaCu_2O_{8+\delta}$ using spectroscopic ellipsometry, but were not able to resolve the structure that we see with the more sensitive TDR spectrometer.

Shaw and Swihart's expression was derived for s-wave superconductors. In a d-wave superconductor the energy gap has momentum dependence and an integral has to be taken over these terms before the data can be cast in their form. A symmetry argument can be given that shows that the contributions of adjacent lobes, where the gaps have opposite signs, have the same value. Both lead to a reduction of the optical conductivity ratio where the absolute value of $\Delta(\omega)$ is large.

In Fig. 1 we show a plot of the measured value of $Re[\sigma_S(\omega)/\sigma_N(\omega)]$ of Tl2212 at 90K vs. photon energy; and the corresponding calculated value derived with the trial coupling function, $G(\omega)$ that gave the best fit to the data. These are compared with the Mattis-Bardeen plot for an energy gap of 6meV at this temperature. Deviations from the Mattis-Bardeen plot occur where $G(\omega)$ is large. The coupling functions in the different regions are found to be: $\lambda_{phonon} = 0.91$, $\lambda_{1.2eV} = 0.114$, and $\lambda_{1.7eV} = 0.294$, giving $\lambda_{total} = 1.318$ and a $T_c$ of 105K.

We had suggested in our earlier work that these electronic excitations [17][18] might be the $d^9$-$d^{10}\underline{L}$ charge transfer excitations, but later turned to the d-to-d transitions of the Cu-ion [25]. A definite assignment was needed.

In 1993, Tanaka and Kotani [28] showed theoretically that the electric-dipole forbidden d-to-d excitations of transition elements could be studied by Resonant Inelastic X-Ray emission spectroscopy (RIXS). In Cu, for example a 2p electron can be excited to an unoccupied 3d level by a photon. This process is dipole allowed. It leaves the atom in an excited state and an electron in any of the 3d levels can then fall to the empty 2p level with the emission of a photon. The difference in energy of the two photons gives the energy of the d-level relative to the previously unoccupied level, and the polarization helps identify which d-level is involved.

Ghiringhelli *et al* [29] have used this technique to study several of the superconducting cuprates. The d-d transitions fall in the region between 1.0 and 2eV, where we had observed structure in the gap function. RIXS data on Tl2212 promised a positive identification.

We obtained several highly reflective samples of Tl2212 films, 2" in diameter, c-axis oriented, 500-700nm thick on lanthanum aluminate substrates with a Tc > 102K manufactured by DuPont, which were similar to those used in the optical study. The RIXS experiment was carried out at the European Synchrotron Radiation Facility in Grenoble, France using the AXES spectrometer and dedicated monochromator [30] [31]. The spectra obtained are shown in the upper panel of Figure 2. The coupling function G(ω) required to fit the TDR data for Tl2212 is shown for comparison in the lower panel.

The 1.2eV peak in G(ω) clearly corresponds to the transition xy to $x^2$-$y^2$, and the stronger, 1.7eV peak to the (xz, yz) to $x^2$-$y^2$ transitions. The data in Figure 1(a) suggests that a small addition to G(ω) corresponding to the $z^2$ term, seen in the RIXS data, also would improve the fit of the calculated response of Figure 1(b) near 2eV.

Our conclusion is that the high superconducting transition temperature of the *optimally* doped cuprates is the result of the combined contributions of a moderately strong phonon interaction, and an excitonic-like interaction [10][11] from the d-to-d excitations of the Cu d-shell. We have shown elsewhere [24] that vertex corrections to the excitonic terms in this case, are small.

**Acknowledgments**

We thank Douglas Scalapino, University of California, Santa Barbara, Ballam Willemsen of Superconductor Technologies, Inc., Santa Barbara, CA., and Jim McCambridge of DuPont Corporation, Wilmington, DA for providing us with high quality, beautiful Tl2212 samples for the RIXS experiments. The RIXS work was done under the AXES (Advanced X-Ray Emission Spectroscopy) contract between the Istituto Nazionale di Fisica della Materia (INFM) from Italy and the European Synchrotron Radiation Facility (ESRF) in Grenoble, France. We acknowledge financial support from the Department of Energy (Grant DEFG03-86ER4525) and Office of Naval Research contract N00014-96-1-0939.

**Figure Captions**

**Figure 1**. Plot of Re[$\sigma_S(\omega)/\sigma_N(\omega)$] vs. $\omega$ for Tl2212 from (a) TDR data and (b) Calculated response using the coupling function $G(\omega)$ shown in Fig. 2, and compared in each case with the Mattis-Bardeen plot.

**Figure 2**. The upper figure shows the RIXS spectra with the de-convoluted d-d spectrum revealed. The lower figure shows the coupling function, $G(\omega)$ discussed in the text that best fit the TDR spectra. This is plotted vs the negative of the energy to allow comparison with the RIXS energy loss spectra.

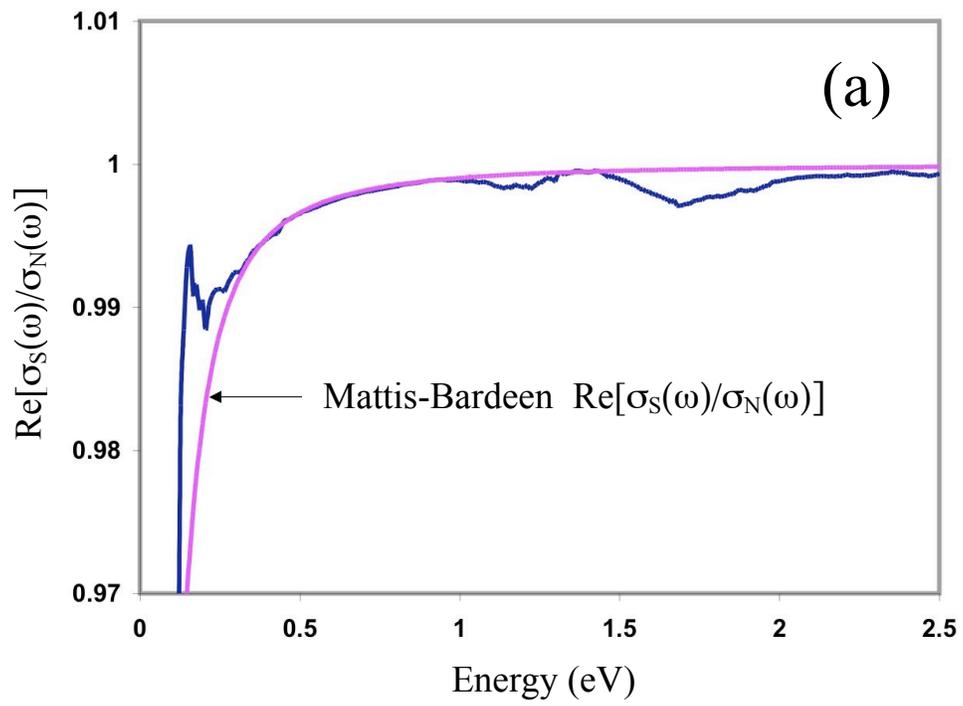

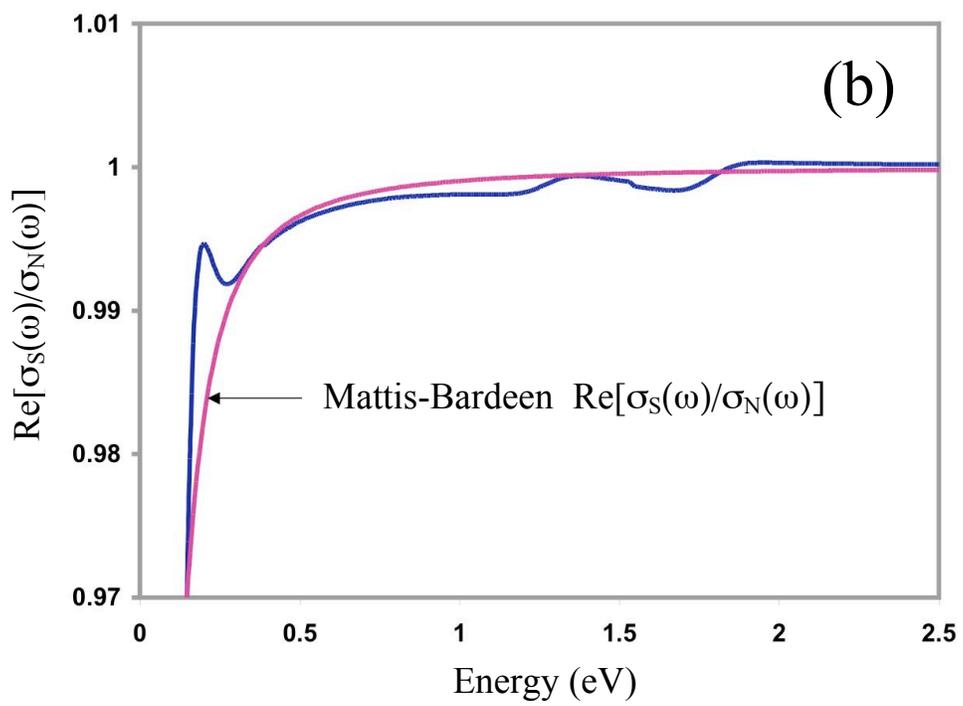

*Figure 1 (W.A. Little, et al.)*

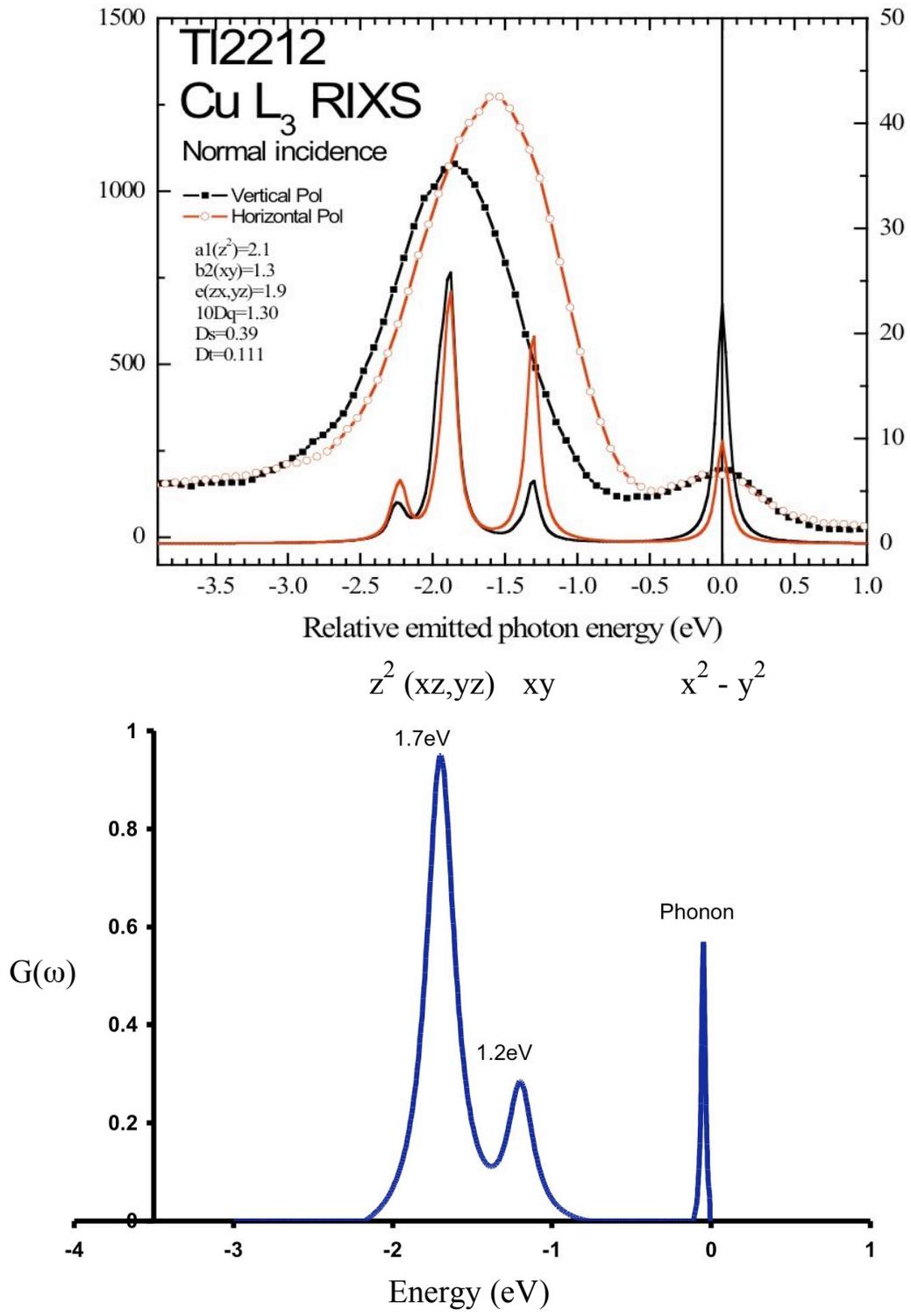

*Figure 2 (W.A. Little, et al.)*